\documentclass[10pt, twocolumn, twoside]{IEEEtran}

\usepackage{cite}
\usepackage{enumerate}
\usepackage{graphicx}
\usepackage[cmex10]{amsmath} 
\usepackage{amssymb}
\usepackage{mathtools}
\usepackage{xspace}
\usepackage{subfigure}
\usepackage[lined,linesnumbered,ruled,commentsnumbered]{algorithm2e} 
\usepackage{colonequals}
\usepackage[ mathscr]{euscript}
\usepackage{fixltx2e}    

\usepackage{epsfig}
\usepackage{epstopdf}

\usepackage{tikz}
\usepackage{pgfplots}

\usetikzlibrary{arrows,shapes,backgrounds,plotmarks,positioning}

\pgfplotsset{compat=newest}                         
\pgfplotsset{plot coordinates/math parser=false}
\newlength\figureheight
\newlength\figurewidth

\newtheorem{theorem}{Theorem}[section]

\newtheorem{proposition}[theorem]{Proposition}

\newtheorem{example}[theorem]{Example}

\newcommand{\C}{\mathcal{K}}         
\newcommand{\Pak}{\mathcal{P}}       
\newcommand{\pv}{\mathbf{p}}         
\newcommand{\F}{\mathbb{F}}          
\newcommand{\Has}{\mathcal{H}}       
\newcommand{\Set}{\mathcal{S}}       
\newcommand{\rv}{\mathbf{r}}         
\newcommand{\Vu}{v_{\alpha}}         

\newcommand{\Xu}{\xi_{\alpha}}

\newcommand{\X}{\mathcal{X}}         
\newcommand{\Y}{\mathcal{Y}}         

\newcommand{\W}{\mathcal{W}_{\mathcal{K}}}
\newcommand{\WS}{\mathcal{W}_{\mathcal{S}}}

\newcommand{\WKX}{\mathcal{W}_{\C\setminus\mathcal{X}}}
\newcommand{\U}{\mathcal{U}}
\newcommand{\M}{\mathcal{M}}
\newcommand{\ZSet}{\mathcal{Z}}

\newcommand{\MAC}{\text{1-MAC}}
\newcommand{\SFM}{\text{SFM}}

\newcommand{\WSO}{\mathcal{W}_{\mathcal{S}}^*}
\newcommand{\WSD}{\mathcal{W}_{\mathcal{S}}^{\circ}}
\newcommand{\AlphaSO}{\alpha_\Set^*}
\newcommand{\AlphaSD}{\alpha_\Set^{\circ}}

\newcommand{\N}{\mathbb{N}_0}

 {\begin{list}{}%
         {\setlength{\leftmargin}{#1}}%
         \item[]%
 }
 {\end{list}}

\begin{document}

\title{Iterative Merging Algorithm for Cooperative Data Exchange}


\author{
\IEEEauthorblockN{Ni~Ding\IEEEauthorrefmark{1}, Rodney~A.~Kennedy\IEEEauthorrefmark{1} and Parastoo~Sadeghi\IEEEauthorrefmark{1}}\\
\IEEEauthorblockA{\IEEEauthorrefmark{1}Research School of Engineering, College of Engineering and Computer Science, the Australian National University (ANU), Canberra, ACT 2601\\
Email: $\{$ni.ding, rodney.kennedy, parastoo.sadeghi$\}$@anu.edu.au}
}


\maketitle

\begin{abstract}
We consider the problem of finding the minimum sum-rate strategy in cooperative data exchange systems that do not allow packet-splitting (NPS-CDE). In an NPS-CDE system, there are a number of geographically close cooperative clients who send packets to help the others recover a packet set. A minimum sum-rate strategy is the strategy that achieves universal recovery (the situation when all the clients recover the whole packet set) with the the minimal sum-rate (the total number of transmissions). We propose an iterative merging (IM) algorithm that recursively merges client sets based on a lower estimate of the minimum sum-rate and updates to the value of the minimum sum-rate. We also show that a minimum sum-rate strategy can be learned by allocating rates for the local recovery in each merged client set in the IM algorithm. We run an experiment to show that the complexity of the IM algorithm is lower than that of the existing deterministic algorithm when the number of clients is lower than $94$.
\end{abstract}


\section{introduction}
\label{sec:Intro}

Due to the growing amount of data exchange over wireless networks and increasing number of mobile clients, the base-station-to-peer (B2P) links are severely overloaded. It is called the `last mile' bottleneck problem in wireless transmissions. Cooperative peer-to-peer (P2P) communications is proposed for solving this problem. The idea is to allow mobile clients to exchange information with each other through P2P links instead of solely relying on the B2P transmissions. If the clients are geographically close to each other, the P2P transmissions could be more reliable and faster than B2P ones.

Consider the situation when a base station wants to deliver a set of packets to a group of clients. Due to the fading effects of wireless channels, after broadcast via B2P links, there may still exist some clients that do not obtain all the packets. However, the clients' knowledge of the packet set may be complementary to each other. Therefore, instead of relying on retransmissions from the base station, the clients can broadcast linear packet combinations of the packets they know via P2P links so as to help the others recover the missing packets. We call this kind of transmission method cooperative data exchange (CDE) and the corresponding system CDE system.

Let the \textit{universal recovery} be the situation that all clients obtain the entire packet set and the \textit{sum-rate} be the total number of linear combinations sent by all clients. In CDE systems, the most commonly addressed problem is to find the minimum-sum rate strategy, the transmission scheme that achieves universal recovery and has the minimum sum-rate. This problem was introduced in \cite{Roua2010}. Randomized and deterministic algorithms for solving this problem have been proposed in \cite{SprintRand2010,AbediniNonMinRank2012} and \cite{MiloDivConq2011,CourtIT2014}, respectively. The idea of the randomized algorithms in \cite{SprintRand2010,AbediniNonMinRank2012} is to choose a client with the maximal or non-minimal rank of the received encoding vectors and let him/her transmit once by using random coefficients from a large Galois field. But, these randomized algorithms repetitively call the rank function, the complexity of which grows with both the number of clients and the number of packets. On the other hand, the authors in \cite{MiloDivConq2011,CourtIT2014} propose deterministic algorithms where the complexity only grows with the number of clients. But, we will show in this paper that the divide-and-conquer (DV) algorithm proposed in \cite{MiloDivConq2011} can not be applied to CDE systems that do not allow packet-splitting (NPS-CDE). Although the deterministic algorithm in \cite{CourtIT2014} can solve NPS-CDE problems, it relies on the submodular function minimization (SFM) algorithm, and the complexity of SFM algorithms is not low.\footnote{There are many algorithms proposed for solving SFM problems. To our knowledge, the algorithm proposed in \cite{Goemans1995} has the lowest complexity $O(K^5\cdot\gamma+K^6)$, where $K$ is the number of clients, and $\gamma$ is the complexity of evaluating a submodular function.}

In this paper, we first use a counter example to show that the DV algorithm in \cite{MiloDivConq2011} can not solve the NPS-CDE problems. We then propose an iterative merging (IM) algorithm, a deterministic algorithm, for finding the minimum sum-rate and corresponding strategy in NPS-CDE systems. The IM algorithm starts with an initial lower estimate of the minimum sum-rate. It recursively merges the clients that require the least number of transmissions for both the local recovery and the recovery of the collectively missing packets.\footnote{Local recovery means the merged clients exchange whatever missing in the packet set that they collectively know so that they share the same common knowledge and can be treated as a single entity.} The IM algorithm updates the estimate of minimum sum-rate whenever it finds that the universal recovery is not achievable. We prove that the minimum sum-rate can be found by starting the IM algorithm. We also show that a minimum sum-rate strategy can be determined by allocating transmission rates for the local recovery in each merged client sets in IM algorithm. We run an experiment to show that the complexity of the IM algorithm is lower than that of the deterministic algorithm proposed in \cite{CourtIT2014} when the number of clients is lower than $94$.

\section{System Model and Problem Statement}
\label{sec:system}

\begin{figure}[tpb]
	\centering
    \scalebox{0.9}{\begin{tikzpicture}

\draw (-2.6,0.3) rectangle (-1.4,-0.3);
\node at (-2,0) {client $1$};
\draw (-2,0.3)--(-2,1)--(-1.7,1)--(-2,0.8)--(-2.3,1)--(-2,1);
\node at (-2,-0.5) {\scriptsize $\Has_1=\{\pv_1,\pv_3,\pv_4,\pv_6,\pv_7\}$};

\draw (-2.6,2.3) rectangle (-1.4,1.7);
\node at (-2,2){client $2$};
\draw (-2,2.3)--(-2,3)--(-1.7,3)--(-2,2.8)--(-2.3,3)--(-2,3);
\node at (-2,1.5) {\scriptsize $\Has_2=\{\pv_1,\pv_2,\pv_3,\pv_5\}$};

\draw (0.4,0.3) rectangle (1.6,-0.3);
\node at (1,0) {client $3$};
\draw (1,0.3)--(1,1)--(1.3,1)--(1,0.8)--(0.7,1)--(1,1);
\node at (1,-0.5) {\scriptsize $\Has_3=\{\pv_1,\pv_5,\pv_6\}$};

\draw (0.4,2.3) rectangle (1.6,1.7);
\node at (1,2){client $4$};
\draw (1,2.3)--(1,3)--(1.3,3)--(1,2.8)--(0.7,3)--(1,3);
\node at (1,1.5) {\scriptsize $\Has_4=\{\pv_3,\pv_5,\pv_6\}$};

\end{tikzpicture}}
	\caption{An example of CDE system where there are four clients that want to obtain eight packets. $\Has_j$ is the has-set of client $j$.}
	\label{fig:CDESystem}
\end{figure}
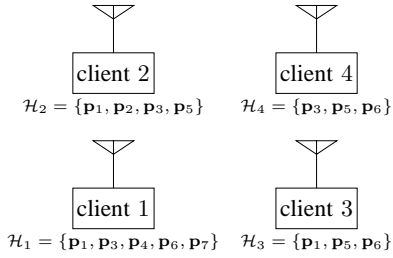

Let $\Pak=\{\pv_1,\dotsc,\pv_L\}$ be the packet set containing $L$ linearly independent packets. Each packet $\pv_i$ belongs to the finite field $\F_q$. The system contains $K$ geographically close clients. Let $\C=\{1,\dotsc,K\}$ be the client set. Each client $j\in\C$ initially obtains $\Has_j\subset\Pak$. Here, $\Has_j$ is called the \textit{has-set} of client $j$. The clients are assumed to collectively know the packet set, i.e., $\cup_{j\in\C}\Has_j=\Pak$. The P2P wireless links between clients are error-free, i.e., information broadcast by any client can be heard losslessly by all other clients. The clients broadcast linear combinations of the packets in their has-sets in order to help each other recover $\Pak$. For example, in the CDE system in Fig.~\ref{fig:CDESystem}, client $1$ broadcasting $\pv_1+\pv_6$ helps client $2$ and client $4$ recover $\pv_6$ and $\pv_1$, respectively. Assume packet-splitting is not allowed. Let $\rv=(r_1,\dotsc,r_K)$ be a transmission strategy with $r_j\in\N$ being the total number of linear combinations transmitted by client $j$. We call $\sum_{j\in\C}r_j$ the \textit{sum-rate} of strategy $\rv$. Let the \textit{universal recovery} be the situation that all clients in $\C$ obtains the entire packet set $\Pak$. The problem is to find a \textit{minimum sum-rate transmission strategy}, a strategy that has the minimum sum-rate among all strategies that achieve universal recovery.

\section{Minimum Sum-rate Strategy}
\label{sec:pre}

In this section, we first clarify the notations, or definitions, that used in this paper and then discuss how to determine the minimum sum-rate and a minimum sum-rate strategy.

Denote $\W$ a \textit{partition} of the client set $\C$.\footnote{A partition $\W$ satisfies $\emptyset\neq\X\subseteq\C$, $\X\cap\X'=\emptyset$ and $\cup_{\X\in\W}\X=\C$ for all $\X,\X'\in\W$.} Let $|\W|$ be the cardinality of $\W$. We call $\W$ a $|\W|$-partition of $\C$. For example, $\W=\{\{1,2\},\{3\},\{4\}\}$ is a $3$-partition of client set $\C=\{1,2,3,4\}$. Let $\Y\subseteq\W$. We call $\Y$ the \textit{$k$-subset} of partition $\W$ if $|\Y|=k$, e.g., $\{\{2\},\{3,4\}\}$ is a $2$-subset of partition $\W=\{\{1\},\{2\},\{3,4\}\}$. Let $\tilde{\Y}=\cup_{\X\in\Y}\X$, e.g., if $\Y=\{\{2\},\{3,4\}\}$, $\tilde{\Y}=\{2,3,4\}$.

For $\Set\subseteq\C$, denote $\rv_{\Set}=\sum_{j\in\Set}r_j$ and $\Has_\Set=\cup_{j\in\Set}\Has_j$. We define the \textit{local recovery} in $\Set$ as the situation such that all clients $j\in\Set$ obtain $\Has_\Set$. For example, in Fig.~\ref{fig:CDESystem}, for $\Set=\{3,4\}$, the problem of local recovery is how to let both client $3$ and client $4$ obtain the packet set $\Has_{\{3,4\}}=\{\pv_1,\pv_3,\pv_5,\pv_6\}$. The minimum sum-rate $\AlphaSO$ for the local recovery in $\Set$ is determined by \cite{Ding2015}
\begin{multline} \label{eq:MinSumRate}
\AlphaSO = \max \Big\{ \Big\lceil \sum_{\X\in\WS}\frac{|\Has_\Set|-|\Has_\X|}{|\WS|-1} \Big\rceil \colon \WS \text{ is a partition} \\ \text{ of } \Set
                \text{ that satisfies } 2\leq{|\WS|}\leq|\Set|\Big\}.
\end{multline}
\eqref{eq:MinSumRate} is based on the condition that $\rv_\X\geq|\Has_{\Set}|-|\Has_{\Set\setminus\X}|$ must be satisfied for all $\X\subset\Set$ for the local recovery in $\Set$ \cite{CourtIT2014}.\footnote{A brief proof of \eqref{eq:MinSumRate} is given in Appendix~\ref{app:MinSumRateProof}.} An equivalent interpretation of \eqref{eq:MinSumRate} is that $\alpha_\Set^*$ is the minimum integer that satisfies
\begin{multline} \label{eq:MinSumRate1}
\AlphaSO \leq \min \{ \sum_{\X\in\WS}(\AlphaSO-|\Has_\Set|+|\Has_\X|) \colon \WS \text{ is a partition} \\ \text{ of } \Set
                \text{ that satisfies } 2\leq{|\WS|}\leq|\Set|\Big\}.
\end{multline}

\begin{proposition} \label{prop1}
In an NPS-CDE system, let $\Set\subseteq\C$, $\Delta{\alpha}_{\Set} = \sum_{\X\in\WSO}(\AlphaSO-|\Has_\Set|+|\Has_\X|) - \AlphaSO$ and $\X'$ be any subset in $\WSO$. A strategy that achieves local recovery in $\Set$ satisfies
\begin{align}
&\rv_{\X}=\AlphaSO-|\Has_\Set|+|\Has_\X|, \forall{\X\in\WSO\setminus\X'},  \nonumber \\
&\rv_{\X'}=\AlphaSO-|\Has_\Set|+|\Has_\X|-\Delta{\alpha}_{\Set}.  \nonumber
\end{align}
\end{proposition}
\begin{IEEEproof}
The constraint conditions $\{\rv_\X\geq|\Has_{\Set}|-|\Has_{\Set\setminus\X}|,\forall{\X}\subset\C, \rv_\C=\AlphaSO\}$ are equivalent to $\{\rv_\X\leq\AlphaSO-|\Has_{\Set}|+|\Has_{\X}|,\forall{\X}\subset\C, \rv_\C=\AlphaSO\}$. Among all these constraints, the tightest ones are $\{\rv_\X\leq\AlphaSO-|\Has_{\Set}|+|\Has_{\X}|,\forall{\X}\subset\WSO,\rv_\C=\AlphaSO\}$, and the excessive rate $\Delta{\alpha}_{\Set}$ can be reduced from any client set in $\WSO$. So, proposition holds.
\end{IEEEproof}

It can be seen that the universal recovery is also the local recovery in $\C$, i.e., \eqref{eq:MinSumRate} and Proposition~\ref{prop1} can be applied for universal recovery by letting $\Set=\C$. We call $\WSO$ the minimum sum-rate partition for the local recovery in $\Set$. Let $\WSD$ be the maximizer of \eqref{eq:MinSumRate}. It should be noted that $\WSD$ and $\WSO$ are not necessarily equal and $\WSD$ can not be used for determining the minimum sum-rate for the local recovery in $\Set$. See the Example~\ref{ex:counter} in the next section.

\section{Errors in Divide-and-conquer Algorithm}
\label{sec:DV}

The authors in \cite{MiloDivConq2011} proposed a divide-and-conquer (DV) algorithm for finding the minimum sum-rate strategy in NPS-CDE systems.\footnote{See details of DV algorithm in Appendix~\ref{app:DV}.} In Theorem 1 in \cite{MiloDivConq2011}, it states that the minimum sum-rate for the local recovery in $\Set$ is given by
\begin{multline} \label{eq:MinSumRateWr}
\AlphaSD = \max \Big\{ \sum_{\X\in\WS}\frac{|\Has_\Set|-|\Has_\X|}{|\WS|-1} \colon \WS \text{ is a partition} \\ \text{ of } \Set
                \text{ that satisfies } 2\leq{|\WS|}\leq|\Set|\Big\}.
\end{multline}
In Lemma 1 in \cite{MiloDivConq2011}, it states that the minimum sum-rate transmission strategy for the local recovery in $\Set$ can be determined by
\begin{equation} \label{eq:lemma}
    \sum_{j\in\X}r_j=\AlphaSD-|\Has_\Set|+|\Has_\X|, \forall\X\in\WSD,
\end{equation}
where $\WSD$ is the maximizer of \eqref{eq:MinSumRateWr}. However, in most of the cases, $\AlphaSD$ is not an integer and Lemma 1 in \cite{MiloDivConq2011} is not correct. For NPS-CDE systems, the minimum sum-rate must be an integer since every client must transmit integer number of times. So, $\AlphaSD$ determined by \eqref{eq:MinSumRateWr} can not necessarily be the minimum sum-rate. It should be round up to a closest integer as in \eqref{eq:MinSumRate}.\footnote{In fact, $\AlphaSO$ determined by \eqref{eq:MinSumRateWr} is the minimum sum-rate for CDE systems that allow packet-splitting (PS-CDE). There is a study in \cite{Ding2015} shows how to determine minimum sum-rate for both PS-CDE and NPS-CDE systems.} One may suggest replacing $\AlphaSD$ by $\AlphaSO$ in \eqref{eq:lemma}. However, if so, Lemma 1 in \cite{MiloDivConq2011} does not hold. See the example below.

\begin{example} \label{ex:counter}
Consider a CDE system in Fig.~\ref{fig:CDESystem}. For the universal recovery, the maximum and the maximizer of \eqref{eq:MinSumRateWr} are $\AlphaSD=13/3$ and $\WSD=\{\{1\},\{2\},\{3\},\{4\}\}$, respectively. The corresponding transmission strategy is $\rv=(7/3,4/3,1/3,1/3)$ by using \eqref{eq:lemma}. This strategy can not be implemented in an NPS-CDE system. Therefore, Lemma 1 in \cite{MiloDivConq2011} is not correct in this case. In addition, if we use $\AlphaSO$ instead of $\AlphaSD$ in \eqref{eq:lemma}, we get $\rv=(3,2,1,1)$, which achieves universal recovery but has a sum-rate greater than $\alpha_\C^*=5$, i.e., it is not a minimum sum-rate strategy. Therefore, Lemma 1 does not hold, either. One may think that reducing $2$ transmissions from any client in strategy $\rv=(3,2,1,1)$ will result in a minimum sum-rate transmission strategy. This is also not true. For example, if we reduce the transmission rate of client 1 by $2$, we get $\rv=(1,2,1,1)$. It has sum-rate equals $\alpha_\C^*=5$. But, the universal recovery is not achievable since constraint $r_1\geq L-|\Has_{\{2,3,4\}}|=2$ is breached. In fact, the correct way is to break $\C$ into $\WSO=\{\{1,2,3\},\{4\}\}$ and determine the individual rates of the clients in $\{1,2,3\}$ for the local recovery in $\{1,2,3\}$.\footnote{See the explanation in Appendix~\ref{app:counter}.} We will show that this can be accomplished by IM algorithm in Example~\ref{ex:1} in the next section.
\end{example}

\section{Iterative Merging Scheduling Method}
\label{sec:idea}
In this section, we propose a greedy scheduling method for the universal recovery in CDE systems. We assume that the clients in CDE system can form coalitions, or groups. A coalition can contain just one client, and each client must appear in no more than one coalition. Any form of coalition in $\C$ can be represented by a partition $\W$, and any $k$-subset $\Y$ of $\W$ contains $k$ coalitions in $\W$. The idea of this scheduling method is to iteratively merge coalitions and check if condition~\eqref{eq:MinSumRate1} holds.

Let $\alpha$ be a lower estimate of $\alpha_\C^*$, e.g., the lower bound on $\alpha_\C^*$ proposed in \cite{SprintRand2010,Ding2015}. At the beginning, we assume that each client forms one coalition, which can be denoted by a $K$-partition $\W=\{\{j\} \colon j\in\C\}$. We start an iterative procedure. In each iteration, we perform two steps:
\begin{enumerate}[1.]
    \item If $\alpha > \sum_{\X\in\W} (\alpha - L + |\Has_{\X}|) $, terminate iteration, increase $\alpha$ by one and start the IM scheduling method (from $K$-partition) again; Otherwise, go to step $2$.
    \item Let $k\in\{2,\cdots,|\W|\}$. We choose $\Y$ as a $k$-subset with the minimum value of $k$ that satisfies the conditions
        \begin{equation}
            \sum_{\X\in\Y}\frac{|\Has_{\tilde{\Y}}|-|\Has_{\X}|}{|\Y|-1} + L -|\Has_{\tilde{\Y}}| < \alpha,   \label{eq:cond1}
        \end{equation}
        \begin{align}
            & \sum_{\X\in\Y}\frac{|\Has_{\tilde{\Y}}|-|\Has_{\X}|}{|\Y|-1} + L-|\Has_{\tilde{\Y}}| \nonumber \\
            &\qquad\qquad\qquad \leq  \sum_{\X\in\Y'}\frac{|\Has_{\tilde{\Y}'}|-|\Has_{\X}|}{|\Y'|-1} + L - |\Has_{\tilde{\Y'}}|, \label{eq:cond2}
        \end{align}
        for all other subsets $\Y'$ such that $|\Y|=|\Y'|$. Achieve local recovery in $\tilde{\Y}$, and update $\W$ by merging all coalitions $\X$ in $\Y$ into one coalition.
\end{enumerate}
The iteration terminates whenever we find that $|\W|=2$ or there is no $k$-subset $\Y$ satisfies conditions~\eqref{eq:cond1} and \eqref{eq:cond2} in step $2$.

Consider step $1$. As discussed in Section~\ref{sec:pre}, if we find that condition
\begin{equation} \label{eq:cond3}
    \alpha \leq \sum_{\X\in\W} (\alpha - L + |\Has_{\X}|)
\end{equation}
does not hold for some partition $\W$, it means $\alpha<\alpha_\C^*$ and universal recovery is not possible with the sum-rate $\alpha$. Therefore, $\alpha$ should be increased.

In step $2$, the interpretations of the conditions \eqref{eq:cond1} and \eqref{eq:cond2} are as follows. Based on Condition~\eqref{eq:cond2}, $\Y$ must be the minimum sum-rate partition for the local recovery of the collectively known packets in $\tilde{\Y}$, i.e., $\Y=\mathcal{W}_{\tilde{\Y}}^*$.\footnote{We will show that this is the case in the proof of Theorem~\ref{theo:main} in Section~\ref{sec:algo}}. So, $\sum_{\X\in\Y}\frac{|\Has_{\tilde{\Y}}|-|\Has_{\X}|}{|\Y|-1}$ incurs the minimum sum-rate for the local recovery in $\tilde{\Y}$. $L-|\Has_{\tilde{\Y}}|$ is the number of collectively missing packets the recovery of which relies on the transmissions in client set $\C\setminus\tilde{\Y}$. If condition~\eqref{eq:cond1} is breached, it means that universal recovery with sum-rate $\alpha$ is not possible if the coalitions in $\Y$ are merged to form one coalition $\tilde{\Y}$. Therefore, it is better for them to work individually than together. Condition~\eqref{eq:cond2} means that $\Y$ require less number of transmissions for the recovery of both collectively known and collectively missing packets than any other $\Y'$ such that $|\Y'|=|\Y|$.

\begin{figure}[tpb]
	\centering
    \subfigure[IM algorithm]{\scalebox{0.65}{\begin{tikzpicture}

\draw  (0,0) circle (0.5);
\node at (0,0) {\Large $\mathbf{1}$};
\node [color=red] at (0.5,-0.4) {$3$};

\draw  (1.5,0) circle (0.5);
\node at (1.5,0) {\Large $\mathbf{2}$};
\node [color=red] at (2,-0.4) {$2$};

\draw  (3,0) circle (0.5);
\node at (3,0) {\Large $\mathbf{3}$};
\node [color=red] at (3.5,-0.4) {$0$};

\draw (1.5,1.5) ellipse (0.8 and 0.4);
\node at (1.5,1.5) {\Large $\mathbf{123}$};

\draw [->,line width=1.5pt] (0.95,1.2)--(0,0.5);
\draw [->,line width=1.5pt] (1.5,1.1)--(1.5,0.5);
\draw [->,line width=1.5pt] (2.05,1.2)--(3,0.5);

\draw  (3.8,1.5) circle (0.5);
\node at (3.8,1.5) {\Large $\mathbf{4}$};
\node [color=red] at (4.3,1.1) {$0$};

\draw [->,line width=1.5pt] (2.3,2.65)--(1.5,1.9);
\draw [->,line width=1.5pt] (3.1,2.65)--(3.8,2);

\draw [dashed] (2.7,3) ellipse (0.8 and 0.4);
\node at (2.7,3) {\Large $\mathbf{1234}$};
\node [color=red] at (3.45,2.6) {$5$};

\coordinate (a) at (5,-0.2);
\coordinate (b) at (5,2.2);
\draw[->, >=latex, red!20!white, line width=7pt]   (a) to node[black]{\rotatebox{90}{bottom-up}} (b) ;

\end{tikzpicture}}}  \quad
    \subfigure[DC algorithm]{\scalebox{0.65}{\begin{tikzpicture}

\draw  (0,0) circle (0.5);
\node at (0,0) {\Large $\mathbf{1}$};
\node [color=red] at (0.5,-0.5) {$\frac{7}{3}$};

\draw  (1.5,0) circle (0.5);
\node at (1.5,0) {\Large $\mathbf{2}$};
\node [color=red] at (2,-0.5) {$\frac{4}{3}$};

\draw  (3,0) circle (0.5);
\node at (3,0) {\Large $\mathbf{3}$};
\node [color=red] at (3.5,-0.5) {$\frac{1}{3}$};

\draw  (4.5,0) circle (0.5);
\node at (4.5,0) {\Large $\mathbf{4}$};
\node [color=red] at (5,-0.5) {$\frac{1}{3}$};

\draw (2.25,2) ellipse (0.8 and 0.4);
\node at (2.25,2) {\Large $\mathbf{1234}$};
\node [color=red] at (3.2,1.7) {$\frac{13}{3}$};

\draw [->,line width=1.5pt] (1.8,1.66)--(0,0.5);
\draw [->,line width=1.5pt] (2.05,1.6)--(1.5,0.5);
\draw [->,line width=1.5pt] (2.35,1.6)--(3,0.5);
\draw [->,line width=1.5pt] (2.65,1.656)--(4.5,0.5);

\coordinate (b) at (6,-0.2);
\coordinate (a) at (6,2.2);
\draw[->, >=latex, blue!20!white, line width=7pt]   (a) to node[black]{\rotatebox{90}{top-down}} (b) ;

\end{tikzpicture}}}
	\caption{The merging process results from iterative merging (IM) algorithm and dividing process results from divide-and-conquer (DC) algorithm when they are applied to find the minimum sum-rate strategy in the CDE system in Fig.~\ref{fig:CDESystem}. Note, final merging to one coalition $\{1,2,3,4\}$ does not happen but is implied in the IM algorithm. In each figure, the minimum sum-rate $\alpha_\C^*$ is shown beside the coalition $\{1,2,3,4\}$, and the rates of clients in minimum sum-rate strategy are shown beside singleton coalitions. Note, the strategy determined by DC algorithm can not be implemented in an NPS-CDE system.}
	\label{fig:bottomup_topdownMain}
\end{figure}
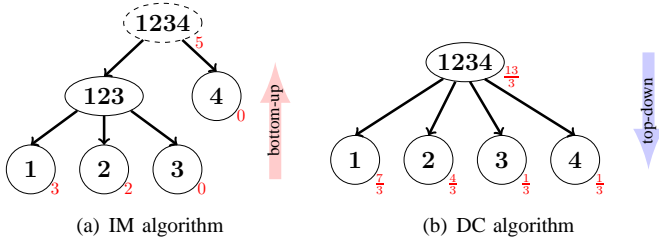

\begin{example} \label{ex:1}
Consider the CDE system in Fig.~\ref{fig:CDESystem} and assume packet-splitting is not allowed. Let $\alpha=5$ and apply IM scheduling method. We get the procedure below.
\begin{itemize}
    \item Assume that each client works individually at the beginning and initiate $\W=\{\{1\},\{2\},\{3\},\{4\}\}$. In this case, we have $\sum_{\X\in\W}(\alpha - L + |\Has_{\X}|)=7>\alpha$, we continue to step $2$ to consider conditions~\eqref{eq:cond1} and \eqref{eq:cond2} to determine which coalitions should be merged. It can be shown that there is no $2$-subset but one $3$-subset $\{1,2,3\}$ satisfies both conditions. Therefore, coalitions $\{1\}$, $\{2\}$ and $\{3\}$ should be merged to form one coalition $\{1,2,3\}$. Consider how to achieve the local recovery in $\{1,2,3\}$. It can be shown that $\alpha_{\{1,2,3\}}^*=\lceil\sum_{\X\in\{\{1\},\{2\},\{3\}\}}\frac{|\Has_{\{1,2,3\}}|-|\Has_\X|}{|\{\{1\},\{2\},\{3\}\}|-1}\rceil=5$ and the minimizer of \eqref{eq:MinSumRate1} is $\mathcal{W}_{\{1,2,3\}}^*=\{\{1\},\{2\},\{3\}\}$. In this case, $\Delta{\alpha}_{\{1,2,3\}} = \sum_{\X\in\mathcal{W}_{\{1,2,3\}}^*}(\alpha_{\{1,2,3\}}^*-|\Has_\Set|+|\Has_\X|) - \alpha_{\{1,2,3\}}^*=1$. According to Proposition~\ref{prop1}, we choose to subtract $\Delta{\alpha}_{\{1,2,3\}}$ from the rate of client $3$ and get rates
        \begin{align}
            &r_1=\alpha_{\{1,2,3\}}^*-|\Has_\{1,2,3\}|+|\Has_1|=3,  \nonumber \\
            &r_2=\alpha_{\{1,2,3\}}^*-|\Has_\{1,2,3\}|+|\Has_2|=2,  \nonumber \\
            &r_3=\alpha_{\{1,2,3\}}^*-|\Has_\{1,2,3\}|+|\Has_3|-\Delta{\alpha}_{\{1,2,3\}}=0.  \nonumber
        \end{align}
    \item For $\W=\{\{1,2,3\},\{4\}\}$, $\sum_{\X\in\W}(\alpha - L + |\Has_{\X}|)=6>\alpha$. However, we do not need to determine the merging candidates since the $2$-partition will be necessarily merged to form coalition $\{1,2,3,4\}$. It is straightforward to see that coalition $\{1,2,3\}$ transmitting $4$ times and coalition $\{4\}$ keeping silent achieve universal recovery. But, there are already $5$ transmissions in $\{1,2,3\}$ when achieving local recovery, which means that client $4$ has recovered the missing packets by listening to the transmissions for the local recovery in $\{1,2,3\}$, i.e., $\rv=(3,2,0,0)$ achieves universal recovery.
\end{itemize}
\end{example}
We will show that $\rv=(3,2,0,0)$ is a minimum sum-rate strategy and $\alpha_\C^*=5$ in Theorem~\ref{theo:main} in the next sections. Note, this procedure also shows a method to determine a minimum strategy: allocate the transmission rates for local recovery in each merged coalition. In Fig.~\ref{fig:bottomup_topdownMain}, we show the merging and dividing processes incurred by IM and DV algorithms, respectively.

\begin{example}
Consider applying the IM scheduling method in the CDE system in Fig.~\ref{fig:CDESystem} with $\alpha=4$. For $\W=\{\{1\},\{2\},\{3\},\{4\}\}$, $\sum_{\X\in\W}(\alpha - L + |\Has_{\X}|)=3<\alpha$. $\alpha$ will be increased to $5$ in the first iteration and IM algorithm starts over again where the same procedure as in Example~\ref{ex:1} is repeated.
\end{example}

\subsection{Iterative Merging Algorithm}
\label{sec:algo}

We describe the IM scheduling method as the IM algorithm in Algorithms 1, 2 and 3. In Algorithm 1, $\Vu$ is defined as
\begin{equation}
    \Vu(\X) = \alpha - L + | \Has_\X|,
\end{equation}
and $\alpha>\sum_{\X\in\W}\Vu(\X)$ is equivalent to the breach of condition~\eqref{eq:cond3}. In Algorithm 2,
\begin{equation}
\Xu(\Y)=\Vu(\tilde{\Y})-\sum_{\X\in\Y}\Vu(\X).
\end{equation}
$\Xu(\Y)<0$ and $\Xu(\Y)\leq\Xu(\Y')$ are equivalent to conditions~\eqref{eq:cond1} and \eqref{eq:cond2}, respectively.

	\begin{algorithm} [t]
	\label{algo:IM}
	\small
	\SetAlgoLined
	\SetKwInOut{Input}{input}\SetKwInOut{Output}{output}
	\SetKwFor{For}{for}{do}{endfor}
    \SetKwRepeat{Repeat}{repeat}{until}
    \SetKwIF{If}{ElseIf}{Else}{if}{then}{else if}{else}{endif}
	\BlankLine
	\Input{$\alpha$, a lower bound on $\alpha_\C^*$}
	\Output{updated $\alpha$, a transmission strategy $\rv$}
	\BlankLine
        $\alpha=\max\{ \alpha,\sum_{j\in\C}\frac{L-|\Has_j|}{K-1}, \sum_{j\in\C} 2L-|\Has_j|-|\Has_{\C\setminus\{j\}}|\}$\; \label{algo:step1}
        initiate a $K$-partition $\W=\{\{j\} \colon j\in\C\}$ and a $K$-dimension transmission strategy $\rv=(0,\cdots,0)$\; \label{algo:start}
        \Repeat{$|\W|=2$ or $\U=\emptyset$}{
            $\U=\text{FindMergeCand}(\W,\alpha)$\;
            $\rv=\text{UpdateRates}(\rv,\U)$\;
            update $\W$ by merging all $\X\in\U$\;
            \If{$\alpha>\sum_{\X\in\W}\Vu(\X)$}{
                $\alpha=\alpha+1$\;
                terminate \lq{repeat}\rq\ loop and go to step~\ref{algo:start}\;
            }
        }
        $\rv=\text{UpdateRates}(\rv,\W)$\;
        $\Delta{r}=\max\{\rv_{\C}-\alpha,0\}$\; \label{algo:steplast1}
        choose $j'\in\tilde{\Y}$ such that $r_{j'}-(L-|\Has_{\C\setminus\{j'\}}|)\geq\Delta{r}$\;
        $r_{j'}=r_{j'}-\Delta{r}$\; \label{algo:steplast2}
	\caption{Iterative Merge (IM)}
	\end{algorithm}

	\begin{algorithm} [t]
	\label{algo:FindMergeCand}
	\small
	\SetAlgoLined
	\SetKwInOut{Input}{input}\SetKwInOut{Output}{output}
	\SetKwFor{For}{for}{do}{endfor}
    \SetKwRepeat{Repeat}{repeat}{until}
    \SetKwIF{If}{ElseIf}{Else}{if}{then}{else if}{else}{endif}
	\BlankLine
	\Input{a partition of client set $\W$, sum-rate $\alpha$}
	\Output{$\U$, a set contains all candidates for merge}
	\BlankLine
        $k=1$\;
        \Repeat{$k=|\W|-1$ or $\U$ is assigned (i.e., $\ZSet\neq\emptyset$)}{
            $k=k+1$\;
            $\ZSet=\{\Y \colon \Xu(\Y)<0, \Y \text{ is a } k \text{-subset of } \W \}$\;
            \If{$\ZSet\neq\emptyset$}{
            $\U=\Y$, where $\Xu(\Y)\leq\Xu(\Y'), \forall \Y'\in\ZSet$\;}
        }
	\caption{FindMergeCand (find merging candidate)}
	\end{algorithm}

\begin{algorithm} [t]
	\label{algo:UpdateRate}
	\small
	\SetAlgoLined
	\SetKwInOut{Input}{input}\SetKwInOut{Output}{output}
	\SetKwFor{For}{for}{do}{endfor}
    \SetKwRepeat{Repeat}{repeat}{until}
    \SetKwIF{If}{ElseIf}{Else}{if}{then}{else if}{else}{endif}
	\BlankLine
	\Input{merge candidate set $\U$, transmission strategy $\rv$}
	\Output{updated transmission strategy $\rv$}
	\BlankLine
        $\alpha_{\tilde{\Y}}=\big\lceil \sum_{\X\in\Y}\frac{|\Has_{\tilde{\Y}}|-|\Has_\X|}{|\Y|-1} \big\rceil$\;
        $\Delta{\alpha}_{\tilde{\Y}}=\sum_{\X\in\Y}(\alpha_{\tilde{\Y}}-|\Has_{\tilde{\Y}}|-|\Has_\X|)-\alpha_{\tilde{\Y}}$\;
        choose $\X'\in\Y$ such that the rates of clients in $\X'$ have not been assigned\;
        \ForAll{$\X\in\U$}{
            $R=\alpha_{\tilde{\Y}}-|\Has_{\tilde{\Y}}|-|\Has_\X|$\;
            \lIf{$\X=\X'$}{$R=R-\Delta{\alpha}_{\tilde{\Y}}$}
            $\Delta{r}=\max\{R-\rv_\X,0\}$\;
            $r_{j'}=r_{j'}+\Delta{r}$, where $j'$ is the client that is randomly chosen in set $\X$\;
         }
	\caption{UpdateRates (update rates)}
	\end{algorithm}

\begin{theorem} \label{theo:main}
The IM algorithm returns $\alpha_\C^*$ and a minimum sum-rate strategy if the input $\alpha\leq\alpha_\C^*$.
\end{theorem}
\begin{IEEEproof}
Consider the Queyranne's algorithm \cite{Queyranne}
\begin{equation}
\M \colonequals \M\cup\{e\}  \nonumber
\end{equation}
where $e=\arg\min\{\Vu(\M\cup\{u\})-\Vu(\{u\}) \colon u\in\C\setminus\M\}$. Let $\W$ be a partition generated by the IM algorithm. For any $\X\in\W$ such that $\X$ is not a singleton, if we start the Queyranne's algorithm with $\M^{(0)}=\Set\subset\X$, we will get $\M^{(|\X|-|\Set|)}=\X$.\footnote{See Appendix~\ref{app:QueyProof} for the proof and examples.} Due to the crossing submodularity of $\Vu$\cite{CourtIT2014}, at any iteration $m\in\{2,\cdots,K-1\}$ of Queyranne's algorithm \cite{Queyranne}
\begin{equation} \label{eq:Quey}
    \Vu(\M^{(m)}) + \Vu(\{j\}) \leq \Vu(\M^{(m)}\setminus\Set) + \Vu(\Set\cup\{j\}),
\end{equation}
for all $j\in\C\setminus\M^{(m)}$ and $\Set$ such that $\emptyset\neq\Set\subseteq\M^{(m-1)}$.\footnote{See the examples in Appendix~\ref{app:MinPart}.} Also, the clients in $\Y$ merges only if $\Xu(\Y)<0$. $\W$ satisfies $\sum_{\X\in\W}\Vu(\X)\leq\sum_{\X\in\W'}\Vu(\X)$ for all other $\W'$ such that $|\W|=|\W'|$. Alternatively speaking, $\W$ generated by the IM algorithm incurs the minimum values of $\sum_{\X\in\W}\Vu(\X)$. So, if $\alpha\leq\sum_{\X\in\W}\Vu(\X)$ holds for all $\W$ in IM algorithm, it means universal recovery is achievable with sum-rate $\alpha$. Since $\alpha$ is increased by $1$ if condition $\alpha\leq\sum_{\X\in\W}\Vu(\X)$ is breached, the output must equal to $\alpha_\C^*$ if the the input $\alpha\leq\alpha_\C^*$.\footnote{Also note that after step~\ref{algo:step1} in IM algorithm, condition~\eqref{eq:cond3} holds for all $2$- and $K$-partitions of $\C$. Therefore, we do not need to check condition~\ref{eq:cond3} for $2$- and $K$-partitions in the rest steps of IM algorithm.}

Consider an NPS-CDE system having $|\U|$ clients and has-sets $\Has_\X,\forall{\X\in\U}$. Based on \eqref{eq:Quey}, we have $\U=\mathcal{W}_{\tilde{\U}}^*$. According to~\eqref{eq:MinSumRate1}, in UpdtateRates algorithm, $\Delta{r}$ for each $\X\in\U$ determines the number of transmissions required from coalition $\X$ for the local recovery in $\U$ in addition to the local recovery in $\X$. The local recovery is achieved in every merged coalition in IM algorithm. Steps~\ref{algo:steplast1} to \ref{algo:steplast2} in IM algorithm are to reduce the excessive rates $\max\{R-\rv_\X,0\}$ from the client $j'$ such that the current rate $r_{j'}$ is greater than $L-|\Has_{\C\setminus\{j'\}}|$, the lower bound on the rate of client $j'$ for universal recovery. Therefore, the output $\rv$ achieves universal recovery and has sum-rate equal to $\alpha_\C^*$.
\end{IEEEproof}

\section{Complexity}
\label{sec:complexity}
The complexity of the IM algorithm depends on two aspects. One is how close the input lower bound $\alpha$ is to $\alpha_\C^*$, since the IM algorithm will be repeated for $\alpha-\alpha_\C^*+1$ times until it updates $\alpha$ to $\alpha_\C^*$. The other is the complexity of FindMergeCand algorithm which may vary with different NPS-CDE systems. For example, if FindMergeCand returns $\U$ containing $2$-subsets, $\beta$ is $O(K^2\cdot\gamma)$. Here, $\gamma$ is the complexity of running the cardinality function $|\Has_\X|$. The authors in \cite{CourtIT2014} also proposed a deterministic algorithm with complexity $O(K\cdot\SFM(K))$ for searching the the minimum sum-rate and minimum sum-rate strategy in CDE systems. Here, $\SFM(K)$ is the complexity of solving a submodular function minimization problem. To our knowledge, the algorithm proposed in \cite{Goemans1995} has the lowest complexity of $\SFM(K)$ is $O(K^5\cdot\gamma+K^6)$. An experiment in \cite{CourtIT2014} shows that the actual runtime by using MATLAB code is $4 \cdot 10^{-3} \cdot K^{1.85}$ seconds on average.

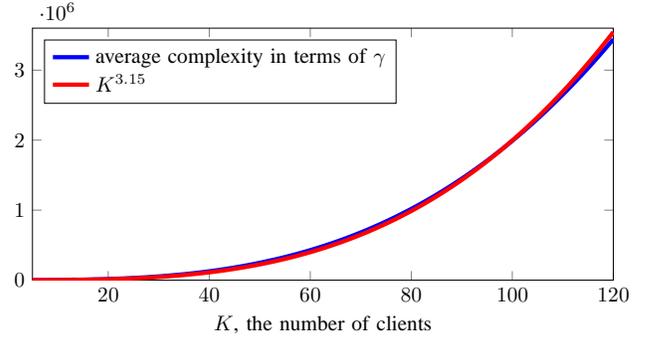
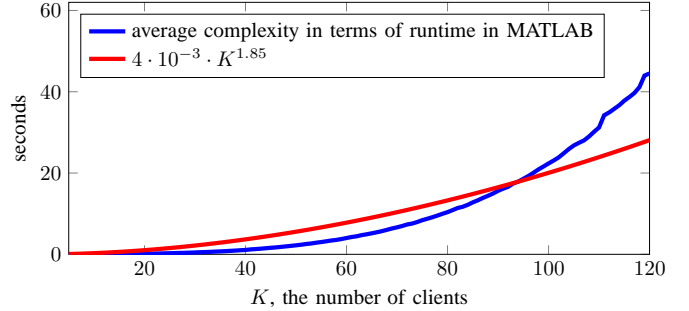
\begin{figure}[tpb]
	\centering
    \subfigure[The average complexity in terms of $\gamma$]{\scalebox{0.8}{
%
%
\begin{tikzpicture}

\begin{axis}[%
width=3.8in,
height=1.65in,
scale only axis,
xmin=5,
xmax=120,
xlabel={$K$, the number of clients},
ymin=0,
ymax=3600000,
legend style={at={(0.02,0.7)},anchor=south west,draw=black,fill=white,legend cell align=left}
]
\addplot [
color=blue,
solid,
line width=2.0pt
]
table[row sep=crcr]{
5 381.3\\
6 455.2\\
7 819\\
8 1158.5\\
9 2266.6\\
10 3120.2\\
11 3653.5\\
12 3685.5\\
13 7252.7\\
14 5934.2\\
15 6723.7\\
16 8740.3\\
17 11205.6\\
18 12226.8\\
19 13391.7\\
20 16042.1\\
21 18595.3\\
22 21415.7\\
23 23862.6\\
24 27119.4\\
25 30677.1\\
26 34903.8\\
27 38699.9\\
28 43175.4\\
29 48003.9\\
30 53205.7\\
31 58679.7\\
32 64958.6\\
33 71049.5\\
34 77519.4\\
35 84609.3\\
36 92133.7\\
37 100007.7\\
38 108375.4\\
39 117209.3\\
40 126494.3\\
41 136254.6\\
42 146492.7\\
43 157288.6\\
44 168563\\
45 180329.3\\
46 192691\\
47 205538.5\\
48 219062.6\\
49 233038\\
50 247745.4\\
51 262985\\
52 279609.8\\
53 295053.1\\
54 312265.4\\
55 329852\\
56 348207.4\\
57 367247.7\\
58 386999.8\\
59 407389\\
60 428616.1\\
61 450377.3\\
62 472991.2\\
63 496265.3\\
64 520377.9\\
65 545178.8\\
66 570808.3\\
67 597257.6\\
68 624380.8\\
69 652391.7\\
70 681263.8\\
71 710954.9\\
72 741452.7\\
73 772850.4\\
74 805137\\
75 838332.9\\
76 872361.3\\
77 907290.4\\
78 943599.5\\
79 979994.4\\
80 1017764.8\\
81 1056584.5\\
82 1096178.1\\
83 1137236.3\\
84 1178572.5\\
85 1221325.5\\
86 1264884.7\\
87 1309675.7\\
88 1355384.9\\
89 1402199.8\\
90 1450091.6\\
91 1499088.7\\
92 1549241.4\\
93 1600250.4\\
94 1652584.7\\
95 1705926.6\\
96 1760534.6\\
97 1816501.4\\
98 1872972.7\\
99 1930994.4\\
100 1990196.7\\
101 2050745.7\\
102 2112215.4\\
103 2175091.3\\
104 2239126.2\\
105 2304478\\
106 2371012.8\\
107 2439042.7\\
108 2507975.4\\
109 2578409.3\\
110 2650156\\
111 2723171.9\\
112 2797674.6\\
113 2873356.6\\
114 2950372.5\\
115 3028778.8\\
116 3108744.8\\
117 3189785.3\\
118 3272555.5\\
119 3356406.6\\
120 3441994.9\\
};
\addlegendentry{average complexity in terms of $\gamma$};

\addplot [
color=red,
solid,
line width=2.0pt
]
table[row sep=crcr]{
5 159.131264443303\\
6 282.602785833083\\
7 459.260216326902\\
8 699.41261145825\\
9 1013.5937051603\\
10 1412.53754462275\\
11 1907.15924823227\\
12 2508.53939105373\\
13 3227.91105889532\\
14 4076.64892617284\\
15 5066.25990721073\\
16 6208.37505642659\\
17 7514.74247731136\\
18 8997.22105860802\\
19 10667.7748976791\\
20 12538.4683013046\\
21 14621.4612765985\\
22 16929.0054416831\\
23 19473.4402987621\\
24 22267.189822343\\
25 25322.7593233244\\
26 28652.7325560035\\
27 32269.7690401649\\
28 36186.6015745563\\
29 40416.033921453\\
30 44970.9386448184\\
31 49864.2550869017\\
32 55108.9874700674\\
33 60718.2031122982\\
34 66705.0307462094\\
35 73082.6589326047\\
36 79864.3345606192\\
37 87063.3614273774\\
38 94693.0988908514\\
39 102766.960590262\\
40 111298.413228944\\
41 120300.975415097\\
42 129788.216556287\\
43 139773.755803949\\
44 150271.261044496\\
45 161294.447933931\\
46 172857.078973137\\
47 184972.962621273\\
48 197655.952444893\\
49 210919.946300633\\
50 224778.885549457\\
51 239246.754300633\\
52 254337.578683727\\
53 270065.426147062\\
54 286444.404781176\\
55 303488.66266594\\
56 321212.387240079\\
57 339629.804691942\\
58 358755.179370422\\
59 378602.813215037\\
60 399187.045204213\\
61 420522.250820889\\
62 442622.841534639\\
63 465503.264299504\\
64 489178.00106685\\
65 513661.568312532\\
66 538968.516577767\\
67 565113.43002307\\
68 592110.925994728\\
69 619975.654603252\\
70 648722.298313317\\
71 678365.571544703\\
72 708920.220283808\\
73 740401.021705284\\
74 772822.783803417\\
75 806200.345032854\\
76 840548.573958322\\
77 875882.368913003\\
78 912216.657665232\\
79 949566.397093205\\
80 987946.572867425\\
81 1027372.19914058\\
82 1067858.31824461\\
83 1109420.00039468\\
84 1152072.34339989\\
85 1195830.47238036\\
86 1240709.53949062\\
87 1286724.72364901\\
88 1333891.23027288\\
89 1382224.29101948\\
90 1431739.16353221\\
91 1482451.13119226\\
92 1534375.50287531\\
93 1587527.61271313\\
94 1641922.81986009\\
95 1697576.5082642\\
96 1754504.08644278\\
97 1812720.98726235\\
98 1872242.66772297\\
99 1933084.60874647\\
100 1995262.31496888\\
101 2058791.31453659\\
102 2123687.15890639\\
103 2189965.42264913\\
104 2257641.70325694\\
105 2326731.62095391\\
106 2397250.81851019\\
107 2469214.9610593\\
108 2542639.7359187\\
109 2617540.85241339\\
110 2693934.04170259\\
111 2771835.05660935\\
112 2851259.67145302\\
113 2932223.68188452\\
114 3014742.90472435\\
115 3098833.17780318\\
116 3184510.35980515\\
117 3271790.3301136\\
118 3360688.98865928\\
119 3451222.25577098\\
120 3543406.07202849\\
};
\addlegendentry{$K^{3.15}$};

\end{axis}
\end{tikzpicture}%
}}  \quad
    \subfigure[The average complexity in terms of runtime in MATLAB]{\scalebox{0.8}{
%
%
\begin{tikzpicture}

\begin{axis}[%
width=3.8in,
height=1.65in,
scale only axis,
xmin=5,
xmax=120,
xlabel={$K$, the number of clients},
ymin=0,
ymax=62,
ylabel={seconds},
legend style={at={(0.02,0.7)},anchor=south west,draw=black,fill=white,legend cell align=left}
]
\addplot [
color=blue,
solid,
line width=2.0pt
]
table[row sep=crcr]{
5 0.0131187086171682\\
6 0.00910306079755701\\
7 0.0122159614038964\\
8 0.0158774533864869\\
9 0.0249396105734777\\
10 0.0330912926325805\\
11 0.0373242578636059\\
12 0.0422316949550604\\
13 0.0654695342059839\\
14 0.0618421537215041\\
15 0.0702522077384198\\
16 0.0869191460895846\\
17 0.104838459416072\\
18 0.115539055250363\\
19 0.126472273389469\\
20 0.15083307523306\\
21 0.177966085322215\\
22 0.192695520767865\\
23 0.2137162032031\\
24 0.240225425684328\\
25 0.270129798778386\\
26 0.305442897050291\\
27 0.348482732638908\\
28 0.37180557895061\\
29 0.412381870070813\\
30 0.455261200495311\\
31 0.515903052103653\\
32 0.552354387674459\\
33 0.605544054245131\\
34 0.671829487186453\\
35 0.716172049944063\\
36 0.781218239327478\\
37 0.851769571698445\\
38 0.920675522373822\\
39 1.00399373193675\\
40 1.07306688932721\\
41 1.20621417915359\\
42 1.29464811725661\\
43 1.38216521073963\\
44 1.49930376319255\\
45 1.60475170149964\\
46 1.70300836244886\\
47 1.82236696968544\\
48 1.95030603749367\\
49 2.07938310110346\\
50 2.21734358368758\\
51 2.35568801684202\\
52 2.53379857794296\\
53 2.66830317333531\\
54 2.85894545360142\\
55 3.02042146718081\\
56 3.20343140539976\\
57 3.39866397638639\\
58 3.54933959535915\\
59 3.7642211741238\\
60 4.01059400494949\\
61 4.25852621423349\\
62 4.43004749045029\\
63 4.72380514667038\\
64 4.92872311908294\\
65 5.17052796750412\\
66 5.42619506463978\\
67 5.70241479219457\\
68 6.02450748128547\\
69 6.36350794194163\\
70 6.63907526445467\\
71 6.95860738752413\\
72 7.33862206452446\\
73 7.56939591857435\\
74 7.94581590779149\\
75 8.40168109311352\\
76 8.75219258446218\\
77 9.12640664648956\\
78 9.53312694367618\\
79 9.97309861603916\\
80 10.3796995531696\\
81 10.8730941181514\\
82 11.3948839695882\\
83 11.6904815336047\\
84 12.17991741395\\
85 12.7825023712019\\
86 13.2839689241233\\
87 13.8758999247132\\
88 14.4018572050427\\
89 14.9553015003625\\
90 15.6156779162161\\
91 16.1423636128725\\
92 16.616305507407\\
93 17.4181922656857\\
94 17.8918267039297\\
95 18.5745405362328\\
96 19.2622934820474\\
97 20.0455415668226\\
98 20.8849810696279\\
99 21.6027072152762\\
100 22.3605795151096\\
101 23.0693576110601\\
102 23.8393109960376\\
103 24.8538132126813\\
104 25.8863423598999\\
105 26.7741743659035\\
106 27.4236545127097\\
107 28.0105290119801\\
108 28.9796522196925\\
109 30.137166378983\\
110 31.2051456742092\\
111 34.2043746939776\\
112 34.9034204776455\\
113 35.7901426524739\\
114 36.6810735824258\\
115 37.8008549307416\\
116 38.6944675980011\\
117 39.6957089244735\\
118 41.1396477338796\\
119 43.9757766101382\\
120 44.5136741826674\\
};
\addlegendentry{average complexity in terms of runtime in MATLAB};

\addplot [
color=red,
solid,
line width=2.0pt
]
table[row sep=crcr]{
5 0.0785515030231765\\
6 0.110062609285003\\
7 0.146383243333551\\
8 0.18740296908104\\
9 0.233028282237256\\
10 0.283178313753655\\
11 0.337781965820163\\
12 0.396775910137057\\
13 0.460103135712875\\
14 0.527711863091345\\
15 0.599554712081939\\
16 0.675588050315722\\
17 0.755771474158619\\
18 0.840067388670937\\
19 0.928440663118492\\
20 1.02085834508735\\
21 1.11728942073295\\
22 1.21770461182807\\
23 1.32207620251038\\
24 1.43037789025542\\
25 1.54258465680002\\
26 1.65867265563969\\
27 1.77861911340493\\
28 1.90240224294519\\
29 2.03000116635518\\
30 2.1613958464974\\
31 2.29656702582691\\
32 2.43549617152557\\
33 2.57816542611574\\
34 2.7245575628541\\
35 2.87465594531444\\
36 3.02844449065582\\
37 3.18590763614576\\
38 3.34703030856899\\
39 3.51179789620243\\
40 3.68019622308039\\
41 3.8522115253094\\
42 4.02783042922303\\
43 4.20703993119278\\
44 4.38982737893356\\
45 4.57618045416136\\
46 4.76608715647701\\
47 4.95953578836443\\
48 5.15651494120401\\
49 5.35701348221238\\
50 5.56102054222956\\
51 5.76852550428245\\
52 5.97951799286081\\
53 6.19398786384858\\
54 6.41192519505864\\
55 6.63332027732425\\
56 6.85816360610494\\
57 7.08644587356824\\
58 7.31815796111251\\
59 7.55329093229891\\
60 7.79183602616365\\
61 8.03378465088388\\
62 8.27912837777311\\
63 8.5278589355839\\
64 8.77996820509732\\
65 9.03544821398071\\
66 9.29429113189622\\
67 9.55648926584444\\
68 9.82203505572837\\
69 10.0909210701242\\
70 10.3631400022465\\
71 10.6386846660959\\
72 10.9175479927791\\
73 11.1997230269906\\
74 11.485202923647\\
75 11.773980944666\\
76 12.0660504558811\\
77 12.3614049240845\\
78 12.6600379141927\\
79 12.9619430865263\\
80 13.2671141941993\\
81 13.5755450806116\\
82 13.8872296770395\\
83 14.2021620003197\\
84 14.5203361506209\\
85 14.8417463093003\\
86 15.1663867368389\\
87 15.4942517708538\\
88 15.8253358241822\\
89 16.1596333830348\\
90 16.4971390052143\\
91 16.8378473183969\\
92 17.1817530184738\\
93 17.5288508679493\\
94 17.8791356943937\\
95 18.232602388948\\
96 18.5892459048791\\
97 18.9490612561814\\
98 19.3120435162254\\
99 19.678187816449\\
100 20.0474893450909\\
101 20.4199433459641\\
102 20.7955451172677\\
103 21.1742900104361\\
104 21.5561734290224\\
105 21.9411908276169\\
106 22.3293377107976\\
107 22.7206096321124\\
108 23.1150021930906\\
109 23.5125110422843\\
110 23.9131318743371\\
111 24.3168604290798\\
112 24.7236924906514\\
113 25.1336238866453\\
114 25.5466504872799\\
115 25.962768204591\\
116 26.3819729916471\\
117 26.8042608417866\\
118 27.2296277878743\\
119 27.6580699015794\\
120 28.0895832926709\\
};
\addlegendentry{$4 \cdot 10^{-3} \cdot K^{1.85}$};

\end{axis}
\end{tikzpicture}%
}}
	\caption{The average complexity over $100$ repetitions in experiment in Section~\ref{sec:complexity}. }
	\label{fig:ComplexityIM}
\end{figure}

We run an experiment to show the actual complexity of the IM algorithm. We set the number of packets $L=50$ and vary the number of clients $K$ from $5$ to $120$. For each value of $K$, we repeat the procedure below for $100$ times.
\begin{itemize}
    \item randomly generate the has-sets $\Has_j$ for all ${j\in\C}$ subject to the condition $\cup_{j\in\C}\Has_j=\Pak$;
    \item set $\alpha$ to be the lower bound on $\alpha_\C^*$ derived in \cite{Ding2015}; run the IM algorithm in MATLAB.
\end{itemize}
In each repetition, we count the actual complexity in terms of $\gamma$ and runtime (including the complexity of using algorithm in \cite{Ding2015} to determine the lower bound on $\alpha_\C^*$). We plot the average complexity over $100$ repetitions in Fig.~\ref{fig:ComplexityIM}. It shows that the average complexity is about $O(K^{3.15}\cdot\gamma)$. The runtime of the IM algorithm is less than that of the deterministic algorithm in \cite{CourtIT2014} when the number of clients is no greater than $94$.

\section{Conclusion}

This paper proposed an IM algorithm that found the minimum sum-rate and a minimum sum-rate strategy in NPS-CDE systems. The IM algorithm started with a sum-rate estimate, a lower bound on the minimum sum-rate. It recursively formed client sets into coalitions and updated the estimate to the value of minimum sum-rate. We proved that a minimum sum-rate strategy could be found by determine individual rate for achieving local recovery in each merged coalitions in the IM algorithm. Based on experiment results, we showed that the complexity of the IM algorithm was lower than the complexity of existing algorithms when the number of clients is below $94$.

\bibliographystyle{ieeetr}
\bibliography{CDE_IMmethodBIB}

\newpage

\appendices

\begin{center}
\Large Additional Pages (Appendices)
\end{center}

\section{}
\label{app:MinSumRateProof}

Eq.~\eqref{eq:MinSumRate} is based on the condition for the local recovery in $\Set$ \cite{CourtIT2014}:
\begin{equation} \label{eq:RateConstr}
\rv_\X\geq|\Has_{\Set}|-|\Has_{\Set\setminus\X}|, \forall\X\subset\Set,
\end{equation}
which states that the total amount of information sent by set $\Set\setminus\X$ must be at least complementary to that collectively missing in set $\X$. Based on \eqref{eq:RateConstr}, we have
\begin{equation} \label{eq:RateConstr1}
    \AlphaSO\geq\sum_{\X\in\WS}(|\Has_\Set|-|\Has_\X|), \forall\WS.
\end{equation}
Likewise, $\rv_{\Set\setminus\X}\geq|\Has_{\Set}|-|\Has_\X|$ should also be satisfied for all $\X\subset\Set$. We have
\begin{equation} \label{eq:RateConstr2}
    \AlphaSO\geq\sum_{\X\in\WS}\frac{|\Has_\Set|-|\Has_\X|}{|\WS|-1}, \forall\X\subset\Set.
\end{equation}
It is shown in \cite{Ding2015} that $\sum_{\X\in\WS}\frac{|\Has_\Set|-|\Has_\X|}{|\WS|-1}\geq\sum_{\X\in\WS}(|\Has_\Set|-|\Has_\X|)$, i.e., constraint~\eqref{eq:RateConstr2} is tighter than \eqref{eq:RateConstr1}. Also, because packet-splitting is not allowed in an NPS-CDE system, $\AlphaSO$ must be an integer, we have \eqref{eq:MinSumRate}.

\section{}
\label{app:DV}
The divide-and-conquer (DV) algorithm iteratively divide non-singleton elements in the current partition by calling a $\MAC$ function. The function $\MAC(\Set)$ returns two outputs: $\AlphaSD$ and $\WSD$, the maximum and maximizer of \eqref{eq:MinSumRateWr}, respectively. The algorithm is shown in Algorithm 4.

In Algorithm 4, $R_\X=\AlphaSD-|\Has_\Set|+|\Has_\X|$ is supposed to determine the sum-rate in $\X$ for achieving local recovery in $\Set$, which is based on Lemma 1 in \cite{MiloDivConq2011} that the minimum sum-rate transmission strategy for the local recovery in $\Set$ satisfies $\rv_\X=\AlphaSD-|\Has_\Set|+|\Has_\X|$ for all $\X\in\WSD$. $\Delta{r}=R_{\Set}-\AlphaSD$ calculates the difference between the sum-rate in $\Set$ that is required for achieving universal recovery in $\C$ and the sum-rate in $\Set$ for achieving local recovery in $\Set$. Based on Theorem 3 in \cite{MiloDivConq2011}, $\Delta{r}\geq{0}$ and it can be added to any set in $\WSD$. The DV algorithm is repetitively called until the rate of each client is determined. It is claimed in \cite{MiloDivConq2011} that $\rv=\text{DV}(\C)$ is the minimum sum-rate strategy for an NPS-CDE system.\footnote{By calling $\text{DV}(\C)$, $\C$ will be assigned to the first input argument $\Set$. $R_\Set$ is not used if $\Set=\C$.} However, the counter example in Section~\ref{sec:DV} shows that it is correct. In fact, one can show that $\rv=\text{DV}(\C)$ is the minimum sum-rate strategy for a PS-CDE system instead of NPS-CDE system.

	\begin{algorithm} [t]
	\label{algo:DV}
	\small
	\SetAlgoLined
	\SetKwInOut{Input}{input}\SetKwInOut{Output}{output}
	\SetKwFor{For}{for}{do}{endfor}
    \SetKwRepeat{Repeat}{repeat}{until}
    \SetKwIF{If}{ElseIf}{Else}{if}{then}{else if}{else}{endif}
	\BlankLine
	\Input{$\Set$, $R_\Set$ if $\Set\neq\C$}
	\Output{$(R_j:j\in\Set)$, where $R_j$ determines transmission rate of client $j$}
	\BlankLine
        initialize $\Delta{r}=0$\;
        $(\AlphaSD,\WSD)=\MAC(\Set)$\;
        \lIf{$\Set\neq\C$}{$\Delta{r}=R_{\Set}-\AlphaSD$}
        \lForAll{$\X\in\WSD$}{$R_{\X}=\AlphaSD-|\Has_{\Set}|+|\Has_{\X}|$}
         randomly choose $\X'\in\WSD$, do $R_{\X'}=R_{\X'}+\Delta{r}$\;
         \ForAll{$\X\in\WSD$ such that $\X$ is not singleton}{$(R_j:j\in\X)=\text{DV}(\X,R_{\X})$\;}

	\caption{DV (divide-and-conquer)}
	\end{algorithm}

\section{}
\label{app:counter}

$\W^\circ=\{\{1\},\{2\},\{3\},\{4\}\}=\{\{j\} \colon j\in\C \}$ implies: Among all constraints $\rv_\X \geq L-\Has_{\C\setminus\X}$ for all $\X\subset\C$ (for the universal recovery), the tightest constrains are $\rv_{\C\setminus\{j\}} \geq L-\Has_{\{j\}}$ for all $j\in\C$.\footnote{Alternatively speaking, a minimum sum-rate strategy must satisfy $r_j \geq L-\Has_{\C\setminus\{j\}}$ and $\rv_{\C\setminus\{j\}} \geq L-\Has_{\{j\}}$ for all $j\in\C$ and $\rv_{\C}=\alpha_\C^*$. } Based on \eqref{eq:MinSumRate}, $\alpha_\C^*$ is an integer such that
\begin{equation}
\alpha_\C^* \geq \sum_{\X\in\W}\frac{L-|\Has_\X|}{|\W|-1}
\end{equation}
for all $\W$ such that $\W$ is a partition of $\C$ that satisfies $2\leq{|\W|}\leq K$. Equivalently,
\begin{equation}
\alpha_\C^* \leq \sum_{\X\in\W} \Vu(\X)
\end{equation}
for all $\W$ such that $2\leq{|\W|}\leq K$, where $\Vu(\X) = \alpha - L + | \Has_\X|$. In the NPS-CDE system in Example~\ref{ex:counter}, we have
\begin{equation}
\sum_{\X\in\W}\frac{L-|\Has_\X|}{|\W|-1} \leq \alpha_\C^*-\frac{1}{|\W|-1}  \nonumber
\end{equation}
 for all $\W$ such that $|\W|>2$ and $L-|\Has_{\X}|+L-|\Has_{\C\setminus\X}|=\alpha_\C^*-1$ for all $\X\in\W^\circ$. One can show that
\begin{multline}
\Vu(\C\setminus\X) = \min \Big\{ \sum_{\Set\in\WKX} \Vu(\Set) \colon \WKX \text{ is a } \\ \text{partition} \text{ of } \C\setminus\X \Big\}
\end{multline}
for all $\X\in\W^\circ=\{\{1\},\{2\},\{3\},\{4\}\}$. For example, let $\X=\{4\}$. Based on $\sum_{j\in\C}\frac{L-|\Has_j|}{3}\leq\alpha_\C^*-\frac{1}{3}$ and $L-|\Has_{\{1,2,3\}}|+L-|\Has_{4}|=\alpha_\C^*-1$, we have
\begin{align}
L-|\Has_1|+L-|\Has_2|+L-|\Has_3| &\leq 3\alpha_\C^*-L+|\Has_4|-1 \nonumber\\
                                 &=2\alpha_\C^*+L-|\Has_{\{1,2,3\}}|,   \nonumber
\end{align}
which is equivalent to $\Vu(\{1,2,3\})\leq\Vu(\{1\})+\Vu(\{2\})+\Vu(\{3\})$. Based on $\frac{L-|\Has_{\{1,2\}}|+L-|\Has_3|+L-|\Has_4|}{2}\leq\alpha_\C^*-\frac{1}{2}$, we can prove that $\Vu(\{1,2,3\})\leq\Vu(\{1,2\})+\Vu(\{3\})$. By this method, one can show that $\Vu(\{1,2,3\})\leq\Vu(\{1,3\})+\Vu(\{2\})$ and $\Vu(\{1,2,3\})\leq\Vu(\{2,3\})+\Vu(\{1\})$. According to Proposition~\ref{prop1}, we can choose any $j\in\C$ to break $\{1,2,3,4\}$ to $\{\C\setminus\{j\},\{j\}\}$ and consider how to achieve local recovery in $\C\setminus\{j\}$. For example, if we break client set to $\{\{1,2,3\},\{4\}\}$, based on Proposition~\ref{prop1}, we get $\rv_{\{1,2,3\}}=5$ and $r_5=0$. Then, consider the individual rates in $\{1,2,3\}$ for the local recovery in $\{1,2,3\}$. We get $r_1=3$, $r_2=2$ and $r_3=0$. One can show that $(3,2,0,0)$ is a minimum sum-rate strategy.

\section{The Proof and Examples of $\M^{(|\X|-|\Set|)}=\X$ by Starting Queyranne's Algorithm with $\emptyset\neq\M^{(0)}=\Set\subset\X$ in the Proof of Theorem~\ref{theo:main}}
\label{app:QueyProof}
Let $\Set\subset\C$ such that $|\Set|\leq K-2$. We have
\begin{equation}
    \Vu(\Set\cup\{u\})-\Vu(\{u\}) = |\cup_{j\in\Set}\Has_j| - |\Has_u \cap (\cup_{j\in\Set}\Has_j)|. \nonumber
\end{equation}
So, $\Vu(\Set\cup\{u\})-\Vu(\{u\}) \leq \Vu(\Set\cup\{u'\})-\Vu(\{u'\})$ is equivalent to
\begin{equation}
    |\Has_u \cap (\cup_{j\in\Set}\Has_j)| \geq |\Has_{u'} \cap (\cup_{j\in\Set}\Has_j)|. \nonumber
\end{equation}
Let $\W$ be the partition of $\C$ that is generated by IM algorithm (at any iteration). For any $\X\in\W$, since the clients $u'\in\C\setminus\X$ is not merged to $\X$, $\Vu(\Set\cup\{u\})-\Vu(\{u\}) \leq \Vu(\Set\cup\{u'\})-\Vu(\{u'\})$, for all $\emptyset\neq\Set\subset\X$, $u\in\X\setminus\Set$ and $u'\in\C\setminus\X$. For example, in Example~\ref{ex:1}, we have $\W=\{\{1,2,3\},\{4\}\}$ in the second iteration. Consider $\X=\{1,2,3\}$. One can show that $|\Has_1\cap\Has_2|\geq|\Has_2\cap\Has_4|$ and $|(\Has_1\cup\Has_2)\cap\Has_3|\geq|(\Has_1\cup\Has_2)\cap\Has_4|$.\footnote{We show two examples of inequalities that can be derived based on $\W=\{\{1,2,3\},\{4\}\}$ and $\X=\{1,2,3\}$. There are in fact many other such inequalities, e.g., $|\Has_1\cap\Has_3|\geq|\Has_3\cap\Has_4|$, $|(\Has_1\cup\Has_3)\cap\Has_2|\geq|(\Has_1\cup\Has_3)\cap\Has_4|$. } They are equivalent to $\Vu(\{2\}\cup\{1\})\leq\Vu(\{2\}\cup\{4\})$ and $\Vu(\{1,2\}\cup\{3\})\leq\Vu(\{1,2\}\cup\{4\})$, respectively.

Therefore, if $\M^{(0)}=\Set$, we will get $\M^{(|\X|-|\Set|)}=\X$ at the $|\X|-|\Set|$th iteration. For example, for $\W=\{\{1,2,3\},\{4\}\}$ in Example~\ref{ex:1} and $\X=\{1,2,3\}$, it can be shown that: If we start the Queyranne's algorithm with $\M^{(0)}=\{1\}$, $\{2\}$ or $\{3\}$, we will get $\M^{(2)}=\{1,2,3\}$; If we start the Queyranne's algorithm with $\M^{(0)}=\{1,2\}$, $\{2,3\}$ or $\{1,3\}$, we will still get $\M^{(1)}=\{1,2,3\}$.

\section{Examples of $\W$ generated by the IM algorithm incurring the minimum values of $\sum_{\X\in\W}\Vu(\X)$}
\label{app:MinPart}

\begin{example}
Consider the NPS-CDE system in Fig.~\ref{fig:CDESystem}. We get $\W=\{\{1,2,3\},\{4\}\}$ in the second iteration of IM algorithm. By applying Queyranne's algorithm with different $\M^{(0)}$, we have the following results:
\begin{itemize}
    \item If $\M^{(0)}=\{1\}$ or $\M^{(0)}=\{3\}$, then $\M^{(1)}=\{1,3\}$ and $\M^{(2)}=\{1,2,3\}$. According to \eqref{eq:Quey}, we have
            \begin{equation}\label{app:MinPart1}
                \Vu(\{1,2,3\})+\Vu(\{4\})\leq\begin{cases}
                                                \Vu(\{2\})+\Vu(\{1,3,4\})  \\
                                                \Vu(\{2,3\})+\Vu(\{1,4\})  \\
                                                \Vu(\{1,2\})+\Vu(\{3,4\})
                                             \end{cases}.
            \end{equation}
    \item If $\M^{(0)}=\{2\}$, then $\M^{(2)}=\{1,2,3\}$. According to \eqref{eq:Quey}, we have
            \begin{equation}\label{app:MinPart2}
                \Vu(\{1,2,3\})+\Vu(\{4\})\leq\Vu(\{1,3\})+\Vu(\{2,4\}).
            \end{equation}
    \item If $\M^{(0)}=\{1,2\}$, then $\M^{(1)}=\{1,2,3\}$. According to \eqref{eq:Quey}, we have
            \begin{equation}\label{app:MinPart3}
                \Vu(\{1,2,3\})+\Vu(\{4\})\leq\Vu(\{3\})+\Vu(\{1,2,4\}).
            \end{equation}
    \item If $\M^{(0)}=\{2,3\}$, then $\M^{(1)}=\{1,2,3\}$. According to \eqref{eq:Quey}, we have
            \begin{equation}\label{app:MinPart4}
                \Vu(\{1,2,3\})+\Vu(\{4\})\leq\Vu(\{1\})+\Vu(\{2,3,4\}).
            \end{equation}
\end{itemize}
$\Vu(\{1,2,3\})+\Vu(\{4\})$ and the right-hand-sides (RHSs) of Eqs.~\eqref{app:MinPart1} to \eqref{app:MinPart4} are the values of $\sum_{\X\in\W}\Vu(\X)$ over all partitions $\W$ of $\{1,2,3,4\}$ such that $|\W|=2$. Therefore, $\Vu(\{1,2,3\})+\Vu(\{4\})$ is the minimum value of $\sum_{\X\in\W}\Vu(\X)$ among all $2$-partitions.
\end{example}

\begin{example}
Consider an NPS-CDE system that contains $5$ clients. They want to obtain a packet set that contains $10$ packets. The has-sets are
\begin{align}
\Has_1&=\{\pv_5,\pv_7,\pv_{10}\},  \nonumber \\
\Has_2&=\{\pv_1,\pv_2,\pv_5,\pv_6,\pv_7,\pv_8,\pv_9\},  \nonumber \\
\Has_3&=\{\pv_1,\pv_3,\pv_5,\pv_6,\pv_7,\pv_8,\pv_9,\pv_{10}\},  \nonumber \\
\Has_4&=\{\pv_1,\pv_3,\pv_4,\pv_5,\pv_6,\pv_7,\pv_8,\pv_9\},  \nonumber \\
\Has_5&=\{\pv_3,\pv_6,\pv_8,\pv_9\}.  \nonumber
\end{align}
We get $\W=\{\{3,4\},\{1\},\{2\},\{5\}\}$ in the first iteration and $\W=\{\{2,3,4\},\{1\},\{5\}\}$ in the second iteration of the IM algorithm. Consider the partition $\W=\{\{2,3,4\},\{1\},\{5\}\}$ in the second iteration. By applying Queyranne's algorithm with different $\M^{(0)}$, we have the following results:
\begin{itemize}
    \item If $\M^{(0)}=\{3\}$ or $\M^{(0)}=\{4\}$, then $\M^{(1)}=\{3,4\}$ and $\M^{(2)}=\{2,3,4\}$. According to \eqref{eq:Quey}, we have
            \begin{align}\label{app:MinPart5}
                &\Vu(\{2,3,4\})+\Vu(\{1\})+\Vu(\{5\})  \nonumber \\
                                            &\leq\begin{cases}
                                                \Vu(\{2\})+\Vu(\{1,3,4\})+\Vu(\{5\})  \\
                                                \Vu(\{2\})+\Vu(\{1\})+\Vu(\{3,4,5\})  \\
                                                \Vu(\{2,3\})+\Vu(\{1,4\})+\Vu(\{5\})  \\
                                                \Vu(\{2,3\})+\Vu(\{1\})+\Vu(\{4,5\})  \\
                                                \Vu(\{2,4\})+\Vu(\{1,3\})+\Vu(\{5\})  \\
                                                \Vu(\{2,4\})+\Vu(\{1\})+\Vu(\{3,5\})
                                             \end{cases}.
            \end{align}
    \item If $\M^{(0)}=\{2\}$, then $\M^{(2)}=\{2,3,4\}$. According to \eqref{eq:Quey}, we have
            \begin{align}\label{app:MinPart6}
                &\Vu(\{2,3,4\})+\Vu(\{1\})+\Vu(\{5\})  \nonumber \\
                                            &\leq\begin{cases}
                                                \Vu(\{3,4\})+\Vu(\{1,2\})+\Vu(\{5\})  \\
                                                \Vu(\{3,4\})+\Vu(\{1\})+\Vu(\{2,5\})
                                             \end{cases}.
            \end{align}
    \item If $\M^{(0)}=\{2,3\}$, then $\M^{(1)}=\{2,3,4\}$. According to \eqref{eq:Quey}, we have
            \begin{align}\label{app:MinPart7}
                &\Vu(\{2,3,4\})+\Vu(\{1\})+\Vu(\{5\})  \nonumber \\
                                            &\leq\begin{cases}
                                                \Vu(\{4\})+\Vu(\{1,2,3\})+\Vu(\{5\})  \\
                                                \Vu(\{4\})+\Vu(\{1\})+\Vu(\{2,3,5\})
                                             \end{cases}.
            \end{align}
    \item If $\M^{(0)}=\{2,4\}$, then $\M^{(1)}=\{2,3,4\}$. According to \eqref{eq:Quey}, we have
            \begin{align}\label{app:MinPart8}
                &\Vu(\{2,3,4\})+\Vu(\{1\})+\Vu(\{5\})  \nonumber \\
                                            &\leq\begin{cases}
                                                \Vu(\{3\})+\Vu(\{1,2,4\})+\Vu(\{5\})  \\
                                                \Vu(\{3\})+\Vu(\{1\})+\Vu(\{2,4,5\})
                                             \end{cases}.
            \end{align}
\end{itemize}
The RHSs of Eqs.~\eqref{app:MinPart5} to \eqref{app:MinPart8} contain a set of minimum values of $\sum_{\X\in\W}\Vu(\X)$. For example, consider the client subset $\{1,3,4,5\}$. Since we $\W=\{\{3,4\},\{1\},\{2\},\{5\}\}$ in the first iteration of the IM algorithm, one can show that $\Vu(\{1,3,4\})+\Vu(\{5\})$ and $\Vu(\{1\})+\Vu(\{3,4,5\})$ incur the minimum values of $\sum_{\X\in\mathcal{W}_{\{1,3,4,5\}}}\Vu(\X)$ over all $2$-partitions of $\{1,3,4,5\}$. So, $\Vu(\{2\})+\Vu(\{1,3,4\})+\Vu(\{5\})$ and $\Vu(\{2\})+\Vu(\{1\})+\Vu(\{3,4,5\})$ at the RHS in \eqref{app:MinPart5} incur the minimum values of $\sum_{\X\in\W}\Vu(\X)$ over all $3$-partitions of $\C=\{1,2,3,4,5\}$ that contains a singleton $\{2\}$. Likewise, one can show that $\Vu(\{3\})+\Vu(\{1,2,4\})+\Vu(\{5\})$ and $\Vu(\{3\})+\Vu(\{1\})+\Vu(\{2,4,5\})$ at the RHS in \eqref{app:MinPart8} incur the minimum values of $\sum_{\X\in\W}\Vu(\X)$ over all $3$-partitions of $\C=\{1,2,3,4,5\}$ that contains a singleton $\{3\}$.

Therefore, $\Vu(\{2,3,4\})+\Vu(\{1\})+\Vu(\{5\})$ is the minimum value of $\sum_{\X\in\W}\Vu(\X)$ over all $3$-partitions of $\C=\{1,2,3,4,5\}$.

\end{example}

\section{Another Example of IM Algorithm}

Consider a $4$-client $8$-packet NPS-CDE system with the has-sets being
\begin{align}
& \Has_1= \{\pv_3,\pv_4,\pv_6,\pv_7,\pv_8\}, \Has_2= \{\pv_1,\pv_4,\pv_7,\pv_8\}  \nonumber \\
& \Has_3=\{\pv_3,\pv_4,\pv_5,\pv_6,\pv_7,\pv_8\}, \Has_4=\{\pv_1,\pv_2,\pv_6\}.  \nonumber
\end{align}
We apply IM algorithm with input $\alpha=6$ and $\alpha=5$. The procedures are shown in Examples~\ref{ex:A1} and \ref{ex:A2}, respectively.

\begin{example} \label{ex:A1}
Let $\alpha=6$. We start IM algorithm. In step~\ref{algo:step1}, we get $\alpha=6$. Then, we initiate $\W=\{\{1\},\{2\},\{3\},\{4\}\}$ and $\rv=(0,0,0,0)$. Then, we have the following procedure.
\begin{itemize}
    \item In the first iteration, we call FindNewPartition to determine $\U$. We first set $k=2$ and consider all $2$-subsets. $\{\{1\},\{2\}\}$, $\{\{1\},\{3\}\}$ and $\{\{2\},\{3\}\}$ are the 2-subsets $\Y$ that satisfy $\Xu(\Y)<0$. Since
        \begin{equation}
            \Xu(\Y) = \begin{cases}
                            -1 & \Y=\{\{1\},\{2\}\} \\
                            -3 & \Y=\{\{1\},\{3\}\} \\
                            -1 & \Y=\{\{2\},\{3\}\}
                      \end{cases},
        \end{equation}
        FindNewPartition returns $\U=\{\{1\},\{3\}\}$ to IM algorithm. We call UpdateRates algorithm, where $r_1$ and $r_3$ in $\rv$ is updated as $r_1=0$ and $r_3=1$, respectively. So, $\rv$ is updated as $\rv=(0,0,1,0)$. We then merge sets $\{1\}$ and $\{3\}$ and update $\W$ as $\W=\{\{1,3\},\{2\},\{4\}\}$. Because $\sum_{\X\in\W}v_6(\X)=v_6(\{1,3\})+v_6(\{2\})+v_6(\{4\})=7>\alpha$ for $\W=\{\{1,3\},\{2\},\{4\}\}$, $\U\neq\emptyset$ and $|\W|>2$, we continue the \lq{repeat}\rq\ loop in IM algorithm.
    \item In the second iteration, FindNewPartition algorithm returns $\U=\{\{1,3\},\{2\}\}$. When calling UpdateRates algorithm, we have $\Delta{r}=2$ for $\{1,3\}$ and $\Delta{r}=1$ for $\{2\}$. we choose $r_1$ to increase by two, and $r_2$ is directly updated as $r_2=1$. The transmission strategy is updated as $\rv=(2,1,1,0)$, and $\W$ is updated as $\W=\{\{1,2,3\},\{4\}\}$. Since $|\W|=2$, the \lq{repeat}\rq\ loop in IM algorithm terminates. Note, in this iteration, $\Delta{r}=2$ for $\{1,3\}$ means that there should be $2$ more transmission from client set, or coalition, $\{1,3\}$ for the local recovery in $\{1,2,3\}$ in addition to the local recovery in $\{1,3\}$. Also, since local recovery has been achieved in $\{1,3\}$, client $1$ and client $3$ have the same has-set at this moment. Therefore, $\Delta{r}=2$ can be completed by any either of them. In this case we added $\Delta{r}$ to client $1$. But, it should be clear that local recovery in $\{1,2,3\}$ can be achieved if we increase $r_3$ by $2$ or increase both $r_1$ and $r_3$ by $1$. Since $|\W|=2$, \lq{repeat}\rq\ loop is terminated.
    \item We call UpdateRates by inputting $\U=\W=\{\{1,2,3\},\{4\}\}$. We get $\Delta{r}=1$ for $\{1,2,3\}$ and $\Delta{r}=1$ for $\{4\}$. We increase $r_2$ by one and set $r_4=1$. The transmission strategy is updated as $\rv=(2,2,1,1)$.
\end{itemize}
Since $\rv_\C=\alpha$, the IM algorithm finally returns $\alpha=6$ and $\rv=(2,2,1,1)$. It can be shown that 6 is the minimum sum-rate and $(2,2,1,1)$ is one of the minimum sum-rate strategies.
\end{example}

\begin{example} \label{ex:A2}
Assume that we apply the IM algorithm to the CDE system in Fig.~\ref{fig:CDESystem} with $\alpha=5$. In step~\ref{algo:step1}, we get $\alpha=5$. However, it can be show that $\W=\{\{1,3\},\{2\},\{4\}\}$ at the end of the first iteration and $v_5(\{1,3\})+v_5(\{2\}) + v_5(\{4\})=4<5$. 'repeat' loop terminates, $\alpha$ is increased to $6$ and the IM algorithm is started over again. With $\alpha=6$, the same procedure as in Example~\ref{ex:A1} is repeated.
\end{example}

We then apply the DV algorithm (in Algorithm 4) to this NPS-CDE system. The procedure and results are shown in the following example.

\begin{example} \label{ex:2}
Consider the NPS-CDE system in Example~\ref{ex:A1}. The first call of $\MAC(\C)$ returns $\alpha_\C^\circ=6$ and $\W^\circ=\{\{1,2,3\},\{4\}\}$, which gives transmission rates $R_{\{1,2,3\}}=\alpha_\C^\circ-L-|\Has_{\{1,2,3\}}|=5$ and $R_{\{4\}}=\alpha_\C^\circ-L-|\Has_{4}|=1$. Then, the problem is to determine the exact rates of clients $1$, $2$ and $3$. To do so, $\MAC(\{1,2,3\})$  is called, which returns $\alpha_{\{1,2,3\}}^\circ=4$ and $\mathcal{W}_{\{1,2,3\}}^\circ=\{\{1,3\},\{2\}\}$. We get $R_{\{1,3\}}=\alpha_{\{1,2,3\}}^\circ-|\Has_{\{1,2,3\}}|+|\Has_{\{1,3\}}|=3$ and $R_{\{2\}}=\alpha_{\{1,2,3\}}^\circ-|\Has_{\{1,2,3\}}|+|\Has_{\{2\}}|=1$ for the local recovery in $\{1,2,3\}$. In this case, there's an excessive rate $\Delta{r}=R_{\{1,2,3\}}-\alpha_{\{1,2,3\}}^\circ=1$ in set $\{1,2,3\}$. $\Delta{r}$ will be added to any client in $\{1,2,3\}$, say, client 2, i.e., $r_2=r_2+1=2$. Then, $\MAC(\{1,3\})$ is called. The results are $\alpha_{\{1,3\}}^\circ=1$ and $\mathcal{W}_{\{1,3\}}^\circ=\{\{1\},\{3\}\}$, which means $R_{\{1\}}=\alpha_{\{1,3\}}^\circ-|\Has_{\{1,3\}}|+|\Has_{\{1\}}|=0$ and $R_{\{3\}}=\alpha_{\{1,3\}}^\circ-|\Has_{\{1,3\}}|+|\Has_{\{3\}}|=1$ are sufficient for the local recovery in $\{1,3\}$. Let the excessive rate $\Delta{r}=2$ be added to client $1$. We finally get the minimum sum-rate strategy $\rv=(2,2,1,1)$.
\end{example}

\begin{figure}[tpb]
	\centering
    \subfigure[IM algorithm]{\scalebox{0.65}{\begin{tikzpicture}

\draw  (0,0) circle (0.5);
\node at (0,0) {\Large $\mathbf{1}$};
\node [color=red] at (0.5,-0.4) {$2$};

\draw  (2,0) circle (0.5);
\node at (2,0) {\Large $\mathbf{3}$};
\node [color=red] at (2.5,-0.4) {$1$};

\draw  (1,1.5) circle (0.5);
\node at (1,1.5) {\Large $\mathbf{13}$};

\draw [->,line width=1.5pt] (0,0.5)--(0.7,1.1);
\draw [->,line width=1.5pt] (2,0.5)--(1.3,1.1);

\draw  (3,1.5) circle (0.5);
\node at (3,1.5) {\Large $\mathbf{2}$};
\node [color=red] at (3.5,1.1) {$2$};

\draw  (2,3) circle (0.5);
\node at (2,3) {\Large $\mathbf{123}$};

\draw [->,line width=1.5pt] (1,2)--(1.7,2.6);
\draw [->,line width=1.5pt] (3,2)--(2.3,2.6);

\draw  (4,3) circle (0.5);
\node at (4,3) {\Large $\mathbf{4}$};
\node [color=red] at (4.5,2.6) {$1$};

\draw [dashed]  (3,4.5) ellipse (0.8 and 0.4);
\node at (3,4.5) {\Large $\mathbf{1234}$};
\node [color=red] at (3.8,4.2) {$6$};

\draw [dashed,->,line width=1.5pt] (2,3.5)--(2.7,4.1);
\draw [dashed,->,line width=1.5pt] (4,3.5)--(3.3,4.1);

\coordinate (a) at (5,0);
\coordinate (b) at (5,4.5);
\draw[->, >=latex, red!20!white, line width=7pt]   (a) to node[black]{\rotatebox{90}{bottom-up}} (b) ;

\end{tikzpicture}}}  \quad
    \subfigure[DC algorithm]{\scalebox{0.65}{\begin{tikzpicture}

\draw  (0,0) circle (0.5);
\node at (0,0) {\Large $\mathbf{1}$};
\node [color=red] at (0.5,-0.4) {$2$};

\draw  (2,0) circle (0.5);
\node at (2,0) {\Large $\mathbf{3}$};
\node [color=red] at (2.5,-0.4) {$1$};

\draw  (1,1.5) circle (0.5);
\node at (1,1.5) {\Large $\mathbf{13}$};

\draw [->,line width=1.5pt] (0.7,1.1)--(0,0.5);
\draw [->,line width=1.5pt] (1.3,1.1)--(2,0.5);

\draw  (3,1.5) circle (0.5);
\node at (3,1.5) {\Large $\mathbf{2}$};
\node [color=red] at (3.5,1.1) {$2$};

\draw  (2,3) circle (0.5);
\node at (2,3) {\Large $\mathbf{123}$};

\draw [->,line width=1.5pt] (1.7,2.6)--(1,2);
\draw [->,line width=1.5pt] (2.3,2.6)--(3,2);

\draw  (4,3) circle (0.5);
\node at (4,3) {\Large $\mathbf{4}$};
\node [color=red] at (4.5,2.6) {$1$};

\draw (3,4.5) ellipse (0.8 and 0.4);
\node at (3,4.5) {\Large $\mathbf{1234}$};
\node [color=red] at (3.8,4.2) {$6$};

\draw [->,line width=1.5pt] (2.7,4.1)--(2,3.5);
\draw [->,line width=1.5pt] (3.3,4.1)--(4,3.5);

\coordinate (b) at (5,0);
\coordinate (a) at (5,4.5);
\draw[->, >=latex, blue!20!white, line width=7pt]   (a) to node[black]{\rotatebox{90}{top-down}} (b) ;

\end{tikzpicture}}}
	\caption{The merging process results from iterative merging (IM) algorithm and dividing process results from divide-and-conquer (DV) algorithm when they are applied to find the minimum sum-rate strategy in the CDE system in Fig.~\ref{fig:CDESystem}. Note, final merging to one coalition $\{1,2,3,4\}$ does not happen but is implied in the IM algorithm. In each figure, the minimum sum-rate $\alpha_\C^*$ is shown beside the coalition $\{1,2,3,4\}$, and the rates of clients in minimum sum-rate strategy are shown beside singleton coalitions. Note, the strategy determined by DC algorithm can not be implemented if packet-splitting is not allowed.}
	\label{fig:bottomup_topdown}
\end{figure}
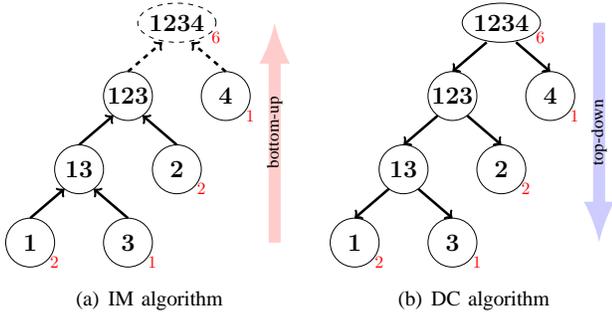

We show the merging and dividing processes resulted from the IM and the DV algorithms in Fig.~\ref{fig:bottomup_topdown}. It can be shown that in this CDE system DV algorithm returns the same results as IM algorithm. And, in this case, the merging process of the IM algorithm is exactly the inverse procedure of the dividing process of the DV algorithm. However, as explained in Section~\ref{sec:DV}, DV algorithm is not applicable to CDE systems that do not allow packet-splitting in general.\footnote{One can show that DV returns a minimum sum-rate strategy for CDE systems that do not allow packet-splitting if all $\WSD$ returns by $\MAC(\C)$ (in Algorithm 4) is a $2$-partition. But, this is not necessarily the case in general.} From Fig.~\ref{fig:bottomup_topdown}, we can see another problem with DV algorithm is that it is a top-down approach which could not be implemented in a decentralized manner. On the contrary, IM algorithm is a bottom-up approach, which allows clients to learn the transmission rates in a minimum sum-rate transmission strategy in a distributed/decentralized way.

\end{document}